\documentclass[aps,prl,twocolumn,a4paper,superscriptaddress,showpacs]{revtex4}

\usepackage{amsmath}
\usepackage{bm}
\usepackage{graphicx}
\usepackage{color}

\begin{document}
    \title{Asymptotic quantum degeneracy at a 2D corner}
    \date{\today}
    \author{S. A. R. Horsley}
    \email{sarh@st-andrews.ac.uk}
    \affiliation{School of Physics and Astronomy, University of St. Andrews, North Haugh, St. Andrews, UK}
    \author{S. Martin--Haugh}
    \affiliation{Department of Physics, The University of Sussex, Brighton, UK}
    \author{M. Babiker}
    \affiliation{Department of Physics, The University of York, Heslington, York, UK}
    \author{M. Al--Amri}
    \affiliation{The National Centre for Mathematics \& Physics, KACST, PO BOX 6086, Riyadh, 11442, Saudi Arabia}

    \begin{abstract}
        In quantum mechanics, asymptotic degeneracy is often considered in the context of a particle in a symmetric
        double--well potential, and is the phenomenon whereby pairs of energy levels come together to form doubly
        degenerate levels in response to an increase in the separation, or depth of the two wells.  Here we highlight
        a new kind of asymptotic degeneracy that can arise when a particle is bound to a surface formed by 
        the intersection of two planes, when the intersection angle is \(>\pi\).  To demonstrate this
        effect we consider the bound states of a charged particle subject its own `image' potential in a
        highly polarizable wedge, as a function of the wedge opening angle, \(\alpha\).
    \end{abstract}
    \maketitle

    \bibliographystyle{unsrt}
%
%
    \par
    Quantum mechanical degeneracy is commonly associated with the invariance of the Hamiltonian under some symmetry group.  \emph{Accidental} degeneracies also exist, that apparently do not arise from any underlying symmetry group (although in such cases the group could just be difficult to find~\cite{leyvraz1997}).  Further to this, there are situations when two energies come so close as to be practically indistinguishable---\emph{asymptotic} degeneracy.  This kind of degeneracy can arise in situations when the probability density is sharply peaked in two (or more) locations in space, but where the probability of tunneling between these locations is negligible.  Asymptotic degeneracy attracted interest in the context of 1D double--well potentials, where such systems were taken as a model for vacuum--vacuum tunneling in gauge theories~\cite{coleman1977,harrell1980,simon1983,shifman1994}.  More recently,  there has been renewed interest in the double--well, as a system where aspects of the physics of Bose--Einstein condensates (BECs) may be observed~\cite{tiecke2003}; for BEC interferometry~\cite{shin2004}; to generate mesoscopic entanglement between binary BECs~\cite{teichmann2007}; and as a model system for a bosonic Josephson junction~\cite{albiez2005}.
    \par
\begin{figure}[h]
        \begin{center}
            \includegraphics[bb=0 0 1875 942,width=0.48\textwidth]{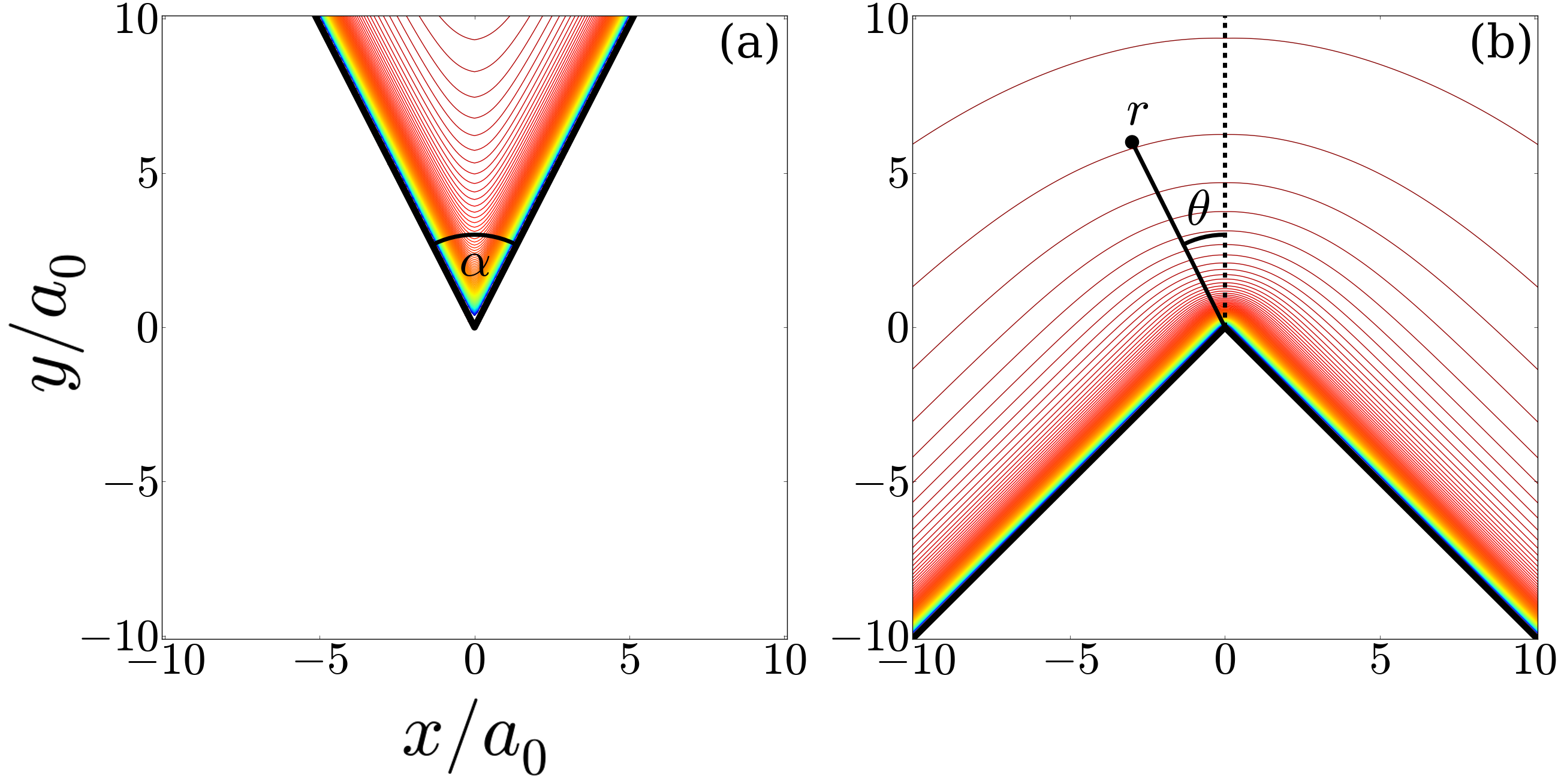}
        \end{center}
        \caption{(Color online) Contour plots for the electrostatic `image' potential given in (\ref{pot_eq}), the dark red lines indicate the highest value of the potential (smallest absolute value), and the yellow, green, and blue lines indicate the potential's lowest values (largest absolute values).  Figure (a) shows a wedge of opening angle, \(\alpha=3\pi/10\), and in (b), \(\alpha=3\pi/2\).\label{figure_1}}
    \end{figure}

    Here we consider a particle confined within a two--dimensional potential formed by the sharp intersection of two planar surfaces; explicitly the system to be investigated consists of a charged particle that is bound---via its own `image' potential, or Coulomb self--energy---to the surface of a highly polarizable material wedge, of opening angle, \(\alpha\) (see figure \ref{figure_1}).  We show that in such cases it is possible---through changing \(\alpha\) continuously across its range, \(0\) to \(2\pi\)---to separate the wavefunction of the particle into two nearly independent parts, as in the case of a double--well, which display an associated degeneracy between pairs of energy levels.  This is different from the usual process of splitting the wave--function, where the splitting is achieved through increasing the strength of the confining potential, or the spatial separation between the wells.  These results may also be relevant in the context of recent work investigating the properties of electronic states bound to liquid helium surfaces (e.g.~\cite{rousseau2009,rees2010}).
%
%
    \par
    Suppose that a particle is subject to a 1D potential, \(V(x)\), that is symmetric around a point, \(x_{0}\), and has two minima, one either side of \(x_{0}\) (a double--well).  It has previously been observed that the eigenvalues of the Hamiltonian for such a system come in pairs that correspond to eigenfunctions that are either symmetric or anti--symmetric around \(x_{0}\).  As the separation between the two potential minima is increased then these pairs of eigenvalues tend to coalesce at a rate that is exponential as a function of distance.  Consequently, for large separations, the system exhibits a degeneracy of symmetric and antisymmetric states~\cite{harrell1980,simon1983}.  The aforementioned splitting between pairs of energy levels is known to be given by,
\(\Delta E=(2\hbar^2/m)\psi_{\text{\tiny{L}}}(x_{0})(d\psi_{\text{\tiny{L}}}/dx)|_{x_{0}}\), where \(\psi_{\text{\tiny{L}}}\) is an eigenstate of just one well, in the absence of the other~\cite[\textsection{50}]{volume3}.
    \par
    We consider a generalization of this system to 2D polar co--ordinates (\(r,\theta\)), where \(\theta\in[-\pi,\pi]\), and the potential is symmetric around the line \(\theta=0\).  The wave--function satisfies the additional boundary condition that it vanishes within the region \(|\theta|\geq\alpha/2\).  If there is no other symmetry than \(V(r,\theta)=V(r,-\theta)\), then there will be two types of states: symmetric, \(\psi^{(+)}(r,\theta)=\psi^{(+)}(r,-\theta)\), and antisymmetric, \(\psi^{(-)}(r,\theta)=-\psi^{(-)}(r,-\theta)\).  Suppose that the wavefunction is confined by the potential so as to be sharply peaked along two channels, \(\theta\sim\pm\alpha/2\), that terminate where they meet, at \(r=0\).  An approximation to the states of this system can be given in terms of single well states, \(\psi^{(\pm)}(r,\theta)\simeq\left[\psi_{\text{\tiny{L}}}(r,\theta)\pm\psi_{\text{\tiny{L}}}(r,-\theta)\right]/\sqrt{2}\) (i.e. where \(\psi_{\text{\tiny{L}}}\) corresponds to a particle confined by the potential along the line \(\theta=+\alpha/2\)).  The difference in energy between the \(\psi^{(\pm)}\) states and the associated single well state, \(\psi_{\text{\tiny{L}}}\), can then be estimated from the current operator,
    \begin{equation}
        -\frac{\hbar^2}{2m}\bm{\nabla}\cdot\left[\psi_{\text{\tiny{L}}}\bm{\nabla}\psi^{(\pm)}-\psi^{(\pm)}\bm{\nabla}\psi_{\text{\tiny{L}}}\right]=\left(E^{(\pm)}-E_{\text{\tiny{L}}}\right)\psi_{\text{\tiny{L}}}\psi^{(\pm)}.\label{eq_1}
    \end{equation}
    Integrating (\ref{eq_1}) over the potential well associated with \(\psi_{\text{\tiny{L}}}\), (i.e. the region, \(\theta>0\)), and applying the divergence theorem we obtain an expression for the energy difference, \(E^{(\pm)}-E_{\text{\tiny{L}}}\),
    \[
         \frac{\hbar^2}{\sqrt{2}m}\int_{0}^{\infty}\left[\psi_{\text{\tiny{L}}}\nabla_{\theta}\psi^{(\pm)}-\psi^{(\pm)}\nabla_{\theta}\psi_{\text{\tiny{L}}}\right]_{\theta=0}dr=E^{(\pm)}-E_{\text{\tiny{L}}},
    \]
    where, \(\nabla_{\theta}=(1/r)\partial/\partial\theta\).  Along \(\theta=0\); \(\psi^{(+)}(r,0)=\sqrt{2}\psi_{\text{\tiny{L}}}(r,0)\); \(\psi^{(-)}(r,0)=0\); \(\nabla_{\theta}\psi^{(+)}(r,0)=0\); and \(\nabla_{\theta}\psi^{(-)}(r,0)=\sqrt{2}\nabla_{\theta}\psi_{\text{\tiny{L}}}(r,0)\).  Therefore the difference between the symmetric and antisymmetric energy levels, \(\Delta E=E^{(-)}-E^{(+)}\), is related to the value of \(\psi_{\text{\tiny{L}}}\) in the following way (c.f.~\cite{harrell1980,simon1983}),
    \begin{equation}
        \Delta E=\left(\frac{2\hbar^2}{m}\right)\int_{0}^{\infty}\psi_{\text{\tiny{L}}}(r,0)(\nabla_{\theta}\psi_{\text{\tiny{L}}})(r,0)\,dr.\label{eq_2}
    \end{equation}
    From (\ref{eq_2}) observe that as \(\psi_{\text{\tiny{L}}}\), or its gradient vanish along the line of symmetry---in this case the region around the corner where the two potential wells meet---the splitting between symmetric and anti--symmetric energy levels goes to zero, and the \(E^{(+)}\) and \(E^{(-)}\) levels become degenerate.
    \par
    As an explicit demonstration of this effect we consider a charge bound within the `image' potential of a highly polarizable material that forms a wedge of opening angle, \(\alpha\), when the particle is restricted from entering the interior of the material (e.g. due to an electronic band--gap within the material that overlaps with the vacuum~\cite{echenique1978,cole1969}---see figure \ref{figure_1}).
%
%
    \par
    For a wedge composed of a material with permittivity \(\epsilon_2\), and an opening of angle \(\alpha\), containing a charged particle, and a second material of permittivity, \(\epsilon_{1}\), the electrostatic potential of the combined wedge+charge system has previously been given in~\cite{scharstein2004}.  This result is a generalization of the case where \(\epsilon_{2}\to\infty\) and \(\epsilon_{1}\to 1\), which was derived some time ago~\cite{macdonald1895,volume8}.  We have obtained the image potential for arbitrary \(\alpha\) through subtracting an integral expression for the potential of a point charge from the expressions given both in~\cite{scharstein2004} and~\cite{volume8}, remembering to divide the result by a factor of two (details will be given in a subsequent publication).  In general the resulting expressions are rather complicated, and for simplicity we consider the case when \(\epsilon_{1}\sim 1\), and \(\Gamma=(\epsilon_{2}-\epsilon_{1})/(\epsilon_{2}+\epsilon_{1})\to 1\), where the potential takes the comparatively compact form,
    \begin{equation}
        \varphi(r,\theta)=\frac{e}{16\pi\epsilon_{0}r}\left[k(\alpha)-f(\theta,\alpha)\right] \label{pot_eq}
    \end{equation}
    with,
    \begin{widetext}
        \begin{align*}
            k(\alpha)&=\left(\frac{2}{\alpha\pi}\right)\int_{0}^{1}\frac{d\eta}{\sqrt{\eta}(1-\eta)^{2}}\left[\frac{(\pi-\alpha)(1-\eta^{(\pi/\alpha+1)})-(\pi+\alpha)(\eta-\eta^{\pi/\alpha})}{1-\eta^{\pi/\alpha}}\right]\\
            f(\theta,\alpha)&=\left(\frac{2}{\alpha}\right)\int_{0}^{1}\frac{d\eta}{\sqrt{\eta}(1-\eta)}\left[\frac{1-\eta^{2\pi/\alpha}}{1+\eta^{2\pi/\alpha}+2\eta^{\pi/\alpha}\cos(2\pi\theta/\alpha)}\right].
        \end{align*}
    \end{widetext}
    This above result is valid for a three dimensional wedge that is infinitely extended along the \(z\) axis, with the assumed symmetry along \(z\) allowing us to neglect this co--ordinate throughout.  The case of arbitrary \(\Gamma\) will be treated elsewhere.  Figure \ref{figure_1} illustrates two contour plots of (\ref{pot_eq}), showing the form of the potential when \(\alpha<\pi\) and \(\alpha>\pi\).  From the figure it can be observed that (\ref{pot_eq}) is an instance of the aforementioned case, where two potential wells extend along the lines \(\theta=\pm\alpha/2\), and meet at \(r=0\).  Note that expression (\ref{pot_eq}) has the correct limit in terms of localized images at the angles, \(\alpha=\pi/n\) (\(n\) integer): for example, at \(\alpha=\pi\), \(k(\alpha)\) vanishes identically, and \(f(\theta,\alpha)=1/\cos(\theta)\) (3.261(2)~\cite{gradshteyn2000}), so that the potential reduces to the known single image result of \(\varphi=-e/(16\pi\epsilon_{0}y)\), where \(y\) is the position on the vertical axis in figure \ref{figure_1}.
%
%
    \begin{figure}[h]
        \begin{center}
            \includegraphics[bb=0 0 545 423,width=0.48\textwidth]{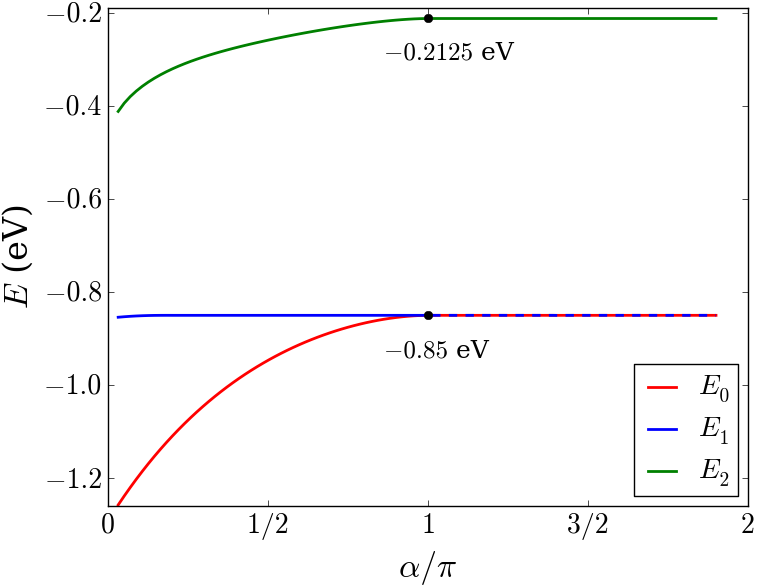}
        \end{center}
        \caption{(Color online) The variation of the upper bounds on the energies obtained from (\ref{hamiltonian_eq}), and (\ref{trial_wavefunctions}), as a function of \(\alpha\).  The circles show the known analytic values of the first two surface states for a particle bound to a planar material with \(\Gamma\to1\).\label{figure_2}}
    \end{figure}
    \par
    Neglecting spin and relativistic effects, the Hamiltonian for a particle of charge \(e\) interacting with the electrostatic potential given in (\ref{pot_eq}) is,
    \begin{equation}
        \hat{H}=-\frac{\hbar^{2}}{2m}\left[\frac{1}{r}\frac{\partial}{\partial r}\left(r\frac{\partial}{\partial r}\right)+\frac{1}{r^{2}}\frac{\partial^{2}}{\partial\theta^{2}}\right]+e\varphi(r,\theta)\label{hamiltonian_eq}
    \end{equation}
    where \(\varphi(r,\theta)\) is given by (\ref{pot_eq}).  It is assumed that the eigenstates of (\ref{hamiltonian_eq}) vanish along the surface, and inside the wedge (\(|\theta|\geq\alpha/2\)).  We find upper bounds for the values of the energies of the first three states using the variational method, with trial wavefunctions that are constructed using the known hydrogenic limit at \(\alpha=\pi\) (e.g. \(\psi_{0}\to N_{0} y e^{-y/4}\)) as a guide,
    \begin{align}
        \psi_{0}(r,\theta)&=N_{0} r^{m_{0}}\cos\left(\frac{\pi\theta}{\alpha}\right)e^{-\gamma_{0}(\theta)r}\nonumber\\
        \psi_{1}(r,\theta)&=N_{1} r^{m_{1}}\sin\left(\frac{2\pi\theta}{\alpha}\right)e^{-\gamma_{1}(\theta)r}\nonumber\\
        \psi_{2}(r,\theta)&=N_{2} r^{m_{2}}\left[a-r\cos\left(\frac{\pi\theta}{\alpha}\right)\right]\cos\left(\frac{\pi\theta}{\alpha}\right)e^{-\gamma_{2}(\theta)r}\label{trial_wavefunctions},
    \end{align}
    where \(\gamma_{i}(\theta)=n_{i}\cos(p_{i}\pi\theta/\alpha)+q_{i}\), and the \(N_{i}\) (\(i=0,1,2\)) are the normalization constants.  The variational parameters are given by, \(\{m_{i},n_{i},p_{i},q_{i}\}\), and are restricted so that they can take only positive values (it is also required that \(p_{i}<1\)), and the constant, \(a\), is chosen so that the orthogonality between \(\psi_{0}\) and \(\psi_{2}\) is guaranteed,
    \[
        a=(m_{0}+m_{2}+2)\frac{I_{2}(\{m_{i},n_{i},p_{i},q_{i}\})}{I_{3}(\{m_{i},n_{i},p_{i},q_{i}\})},
    \]
    where,
    \[
        I_{n}=\int_{0}^{\pi/2}\frac{\cos^{n}(x)dx}{\left[\gamma_{0}(\alpha x/\pi)+\gamma_{2}(\alpha x/\pi)\right]^{m_{0}+m_{2}+n}}.
    \]
    At \(\alpha=\pi\), we expect \(\psi_{2}\) to approach the first, rather than the second excited state.  At this angle, the potential possesses translational symmetry, \(V(x+\delta x)=V(x)\), rather than simply the reflection symmetry assumed for the other angles.  Therefore the eigenstates should satisfy, \(\psi(x+\delta x)\to\psi(x)e^{ik_{x}\delta x}\), which is not possible in the antisymmetric (\(\psi_{1}\)) case.  The completeness of the eigenstates of the Hamiltonian guarantees the existence of the antisymmetric state for all other \(\alpha\).
    \begin{figure}[h]
        \begin{center}
            \includegraphics[bb=0 0 1629 1128,width=0.48\textwidth]{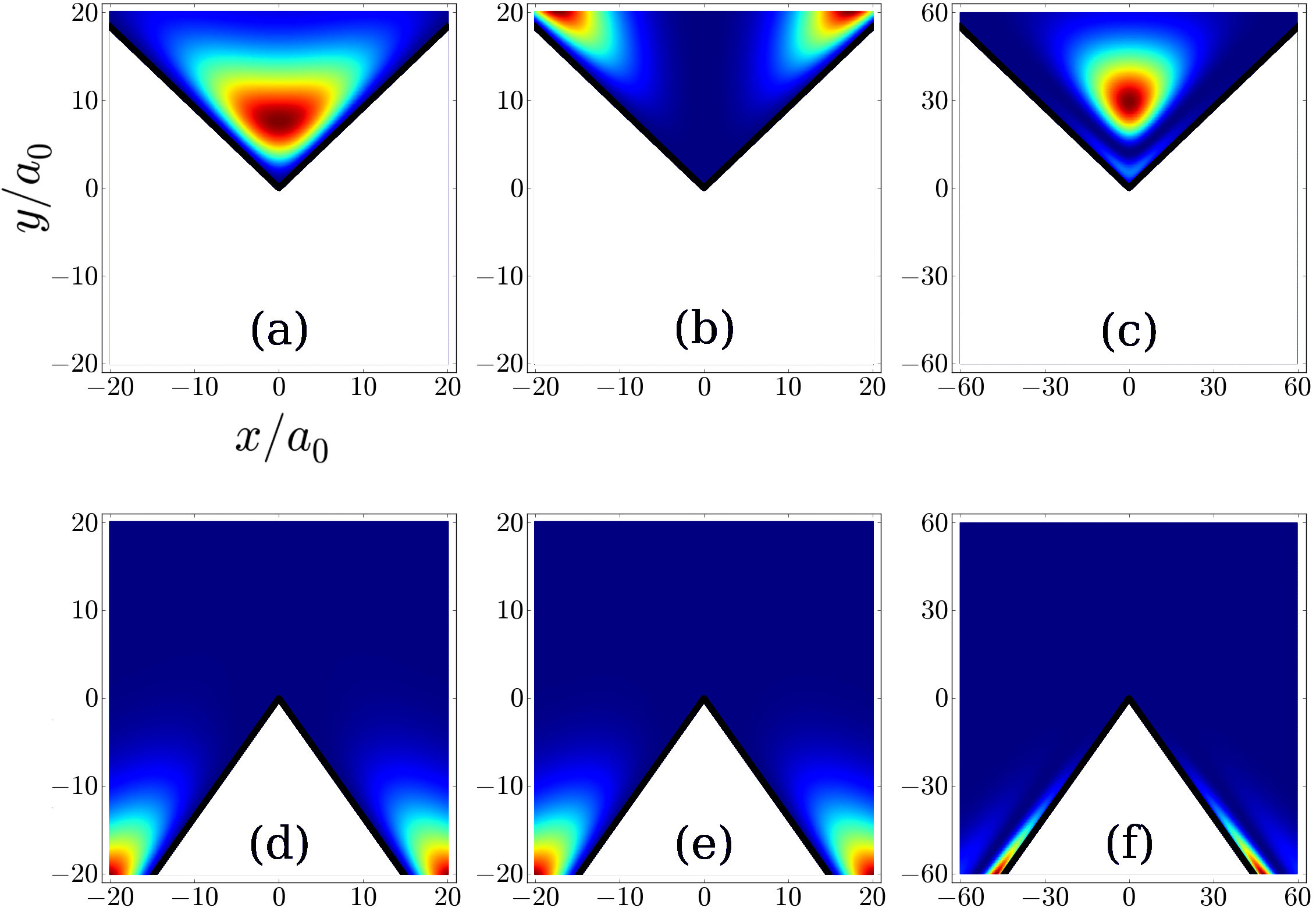}
        \end{center}
        \caption{(Color online) The probability distributions obtained from minimizing \(\langle\psi_{i}|\hat{H}|\psi_{i}\rangle\).  Figures (a)-(c) show \(|\psi_{i}|^{2}\) for \(i=1,2,3\) with \(\alpha= 1.657\;\text{rad}\), while figures (d)-(f) show the corresponding distributions when \(\alpha= 5.011\;\text{rad}\).  Note that the scale of figures (c) and (f) is 3:1 compared to the remaining plots.\label{figure_3}}
    \end{figure}
    \par
    To carry out the variation of the energy we must calculate the integrals of \(\psi_{i}\hat{H}\psi_{i}\) and \(|\psi_{i}|^{2}\) over the interior of the wedge.  It is possible to perform the integration with respect to the radial co--ordinate analytically and write the results in terms of gamma functions (not to be confused with the material contrast parameter, \(\Gamma\), which is here set to \(1\)) multiplying functions of \(\theta\).  For example, the normalization of the state, \(\psi_{0}\), is given by,
    \[
        N_{0}=\sqrt{\frac{2^{2m_{0}+1}\pi/\alpha}{\Gamma(2m_{0}+2)I_{2}(m_{2}=m_{0},n_{2}=p_{2}=q_{2}=0)}},
    \]
    and the associated variational ground state energy is,
    \begin{align*}
        E_{0}=\langle\psi_{0}|\hat{H}|\psi_{0}\rangle=-\frac{\hbar^{2}\alpha N_{0}^{2}\Gamma(2m_{0}+2)}{2^{2m_{0}+2}\pi m}\int_{0}^{\pi/2}\frac{\cos^{2}(x)}{\gamma_{0}^{2m_{0}+2}}\\
        \times\bigg\{\frac{A \gamma_{0}^{2}}{m_{0}(m_{0}+1/2)}-\frac{B \gamma_{0}}{(m_{0}+1/2)}+C\bigg\}\,dx,
    \end{align*}
    where; \(A=m_{0}^{2}-\left(\frac{\pi}{\alpha}\right)^{2}\); \(B=(2m_{0}+1)\gamma_{0}+\left(\frac{\pi}{\alpha}\right)^{2}\left[2\tan{(x)}\gamma_{0}^{\prime}+\gamma_{0}^{\prime\prime}\right]+(me^{2}/8\pi\epsilon_{0}\hbar^{2})g\); \(C=\gamma_{0}^2+\left(\frac{\pi}{\alpha}\right)^{2}{\gamma_{0}^{\prime}}^{2}\); \(g=k(\alpha)-f(\alpha x/\pi,\alpha)\); and \(\gamma_{0}\) and its derivatives are evaluated at \(\alpha x/\pi\).  Similar expressions hold for the quantities relating to the first and second excited states.  The values of these integrals were minimized with respect to the parameters, \(\{m_{i},n_{i},p_{i},q_{i}\}\), using a simplex routine~\cite{scipy}.  The same results were also obtained from a second program based on a different numerical library~\cite{gsl}.  Figure \ref{figure_2} illustrates the resulting upper bounds on the energies, \(E_{0}\), \(E_{1}\), and \(E_{2}\), as a function of the opening angle of the wedge, \(\alpha\).  When \(\alpha<\pi\), the three energies take distinct values that increase at differing rates, as \(\alpha\) increases.  However, as \(\alpha\) approaches \(\pi\), and beyond, \(E_{0}\) and \(E_{1}\) tend to the same value of \(-0.85\;\text{eV}\), remaining degenerate until the wedge becomes a half plane (\(\alpha=2\pi\)).  We interpret this as an asymptotic degeneracy, as outlined above, and expect higher energy levels to pair up in a similar way.
    \par
    In conclusion we have shown that a corner can isolate two parts of the wave--function so that the system behaves as two separate wells, with effectively degenerate symmetric--antisymmetric pairs of states.  This is in contrast to the typical double--well system, where the analogous splitting of the states is directly related to the physical separation, or strength of the potential wells.\\  
    \par
    To explain the origin of the degeneracy we use the distributions associated with the energies of figure \ref{figure_2} as a guide.  Figure \ref{figure_3} (a)-(c) shows these states in the cases when \(\alpha<\pi\), and (d)-(f) when \(\alpha>\pi\).  Although there is no guarantee that these distributions will approach the exact probability distribution as the energy is minimized, it is clear that the minimization of the energy beyond \(\pi\) is achieved through a reduction in the probability density close to the line \(\theta=0\), so that the system behaves as though the problem involved two independent half planes.  Physically this can be understood as a reduction in the average kinetic energy coming from the \(-\psi(\hbar^{2}/2mr^{2})(\partial^{2}\psi/\partial\theta^{2})\) term in the energy density.
    \section{Acknowledgments}
    One of us (S.A.R.H) wishes to acknowledge financial support from the EPSRC and the National Centre for Mathematics \& Physics at KACST, Saudi Arabia. S. M--H wishes to thank Gabriel Barton for useful discussions.

    \bibliography{refs}
\end{document}